 \documentclass[nohyper,12pt,letterpaper]{JHEP3}
 \usepackage{epsfig}
 
\usepackage{graphicx}
\usepackage{epstopdf}
\title{Regulating Eternal Inflation II\\ The Great Divide}
\author{A.\,Aguirre\\
Department of Physics and SCIPP\\
University of California, Santa Cruz, CA 95064\\
E-mail: \email{aguirre@scipp.ucsc.edu}}

\author{T.\,Banks\\
Department of Physics and SCIPP\\
 University of California, Santa Cruz, CA 95064\\
E-mail: \email{banks@scipp.ucsc.edu}\\
{\it and}\\
Department of Physics and NHETC, Rutgers University\\
Piscataway, NJ 08540}

\author{M.\,Johnson\\
Department of Physics and SCIPP\\
University of California, Santa Cruz, CA 95064\\
E-mail: \email{mjohnson@physics.ucsc.edu}}

\abstract{ In a previous paper, two of the authors presented a "regulated" picture of eternal inflation. This picture both suggested and drew support from a conjectured discontinuity in the amplitude for tunneling from positive to negative vacuum energy, as the positive vacuum energy was sent to zero; analytic and numerical arguments supporting this conjecture were given.  Here we show that this conjecture is false, but in an interesting way.
There are no cases where tunneling amplitudes are discontinuous at vanishing cosmological constant; rather, the space of potentials separates into two regions. In one region decay is strongly suppressed, and the proposed picture of eternal inflation remains viable; sending the (false) vacuum energy to zero in this
region results in an absolutely stable asymptotically flat space. In the other region, we argue that the space-time at vanishing cosmological constant is unstable, but {\em not} asymptotically Minkowski. The consequences of our results for theories of supersymmetry breaking are unchanged.}

 \received{} \accepted{}

\begin{document}
 %%%%%%%%%%%%%%%%%%%%%%%%%%%%%%%%%%%%%%%%%%%%%%%%%%%%%%%%%%%%%%%%%%% %%%%%%%%%
% Table of contents automatic !!! %
 %%%%%%%%%%%%%%%%%%%%%%%%%%%%%%%%%%%%%%%%%%%%%%%%%%%%%%%%%%%%%%%%%%%
%%%%%%%%%

\section{\bf Introduction}

The possibility that the universe inflates eternally, to create an
infinite and complex mixture of causally disconnected inflating and
non-inflating regions, is one of the most interesting and perplexing
ideas to emerge in cosmology.  In a recent paper~\cite{etinf}, two of
us (TB and MJ) presented a picture of a large class of eternal
inflation models that greatly simplifies their analysis by viewing
the eternally inflating universe as a finite system comprised of the
causal diamond of a single observer.

This picture, which has consequences for the Landscape idea as well
as for models of low-energy supersymmetry breaking, both suggested
and gained support from an interesting new result in the dynamics of
true-vacuum bubble nucleation as described by Euclidean instanton
techniques.  In particular, it was found that in a certain class of
potentials, the instanton action for a transition from positive (false)
to negative (true) vacuum energy did {\em not} tend to infinity as the false vacuum energy $V_F$ was
reduced to zero, as would be required to give a
finite nucleation probability\footnote{As $V_F\rightarrow 0$, the
required background subtraction becomes infinite, requiring an
infinite instanton action to cancel it and leave a finite decay
probability.} and hence accord with intuition regarding the decay of
Minkowski space to a negative vacuum (``big crunch'') space.  This
result was supported by general analytic arguments, as well as numerical
results for $\epsilon\sim1$, where $\epsilon$ controls the scale in
field value over which the potential varies.  On the basis of these
results it was conjectured that
\begin{enumerate}
\item The same behavior holds at $\epsilon \ll 1$, and
\item for $V_F \equiv 0$, a second (non-compact) instanton, like the
one found in the absence of gravity, exists which allows much faster
decay, so that
\item for all $\epsilon$ there is a discontinuity in the decay rate
as $V_F\rightarrow 0$.
\end{enumerate}

\EPSFIGURE{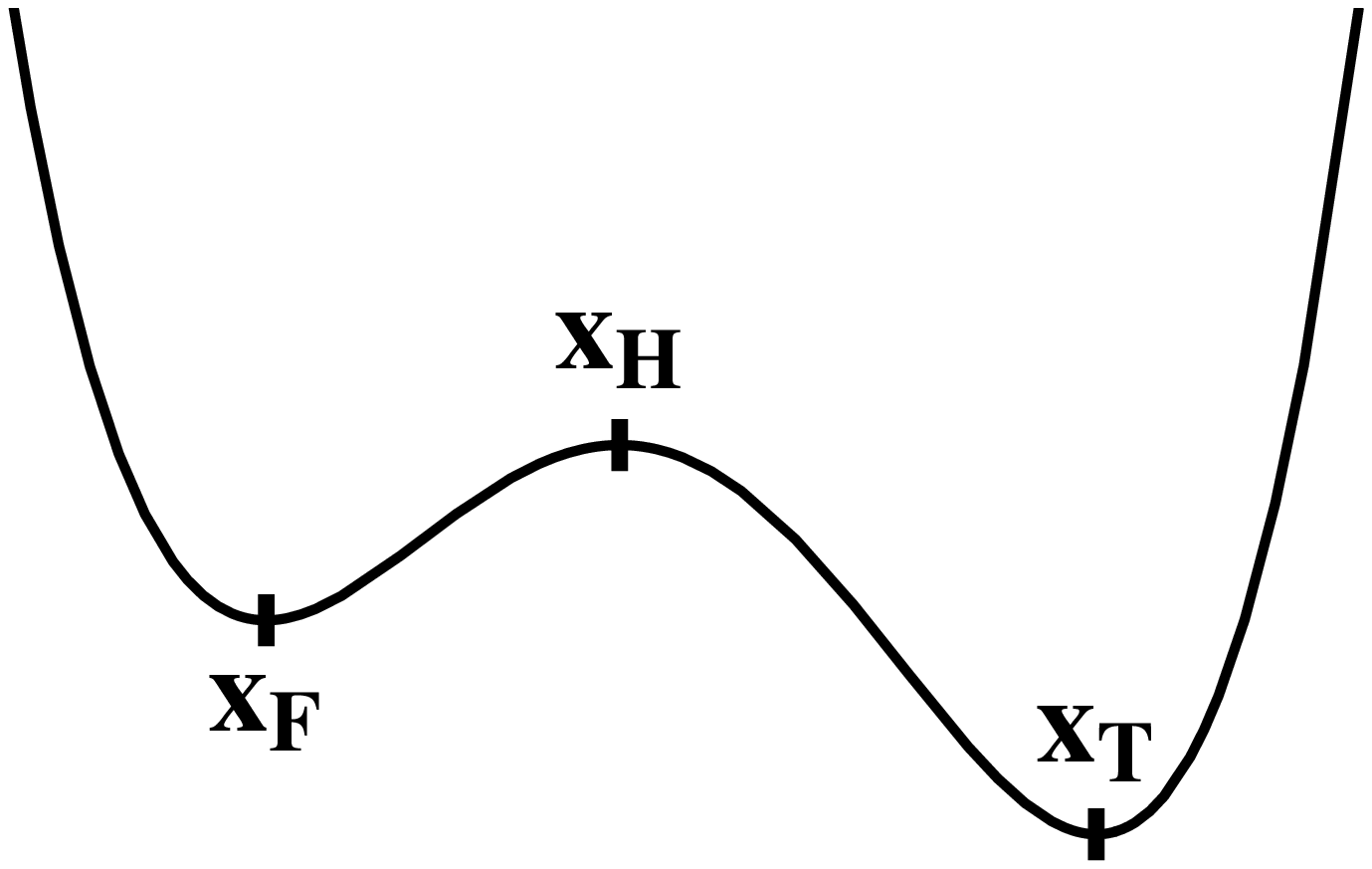,width=6cm}{The potential $V(\phi)$, with the true vacuum $x_T$, the false vacuum $x_F$ and
the ``Hawking-Moss" point $x_H$ labeled.\label{vofx}}

In this paper, we will demonstrate that while the specific
calculations presented in~~\cite{etinf} are correct, the above
conjecture is not\footnote{R. Bousso, B. Freivogel and M. Lippert,
have discovered this fact independently.  Their paper on this
subject will appear simultaneously.}. Instead we find that the space
of potentials is partitioned by a {\it Great Divide}, into one class
where Minkowski space is unstable, and a second class where the
tunneling rate is indeed suppressed -- as argued in~\cite {etinf} --
by the factor $e^{-\pi (RM_P)^2}$ (where $R$ is the de Sitter radius
corresponding to the false vacuum), and hence vanishes at $V_F=0$. The stability, for some potentials, of a seemingly metastable
Minkowski vacuum was noted long ago by Coleman and De
Luccia~\cite{cdl} in the thin-wall limit and subsequently discussed
by several authors~\cite{samuel,heretics} outside of that limit.

In Sections 2-4 we will review the instanton formalism, give
approximate analytic solutions, then examine the behavior of the
instanton solutions in the limit where $V_F\rightarrow 0$, using both
analytic and numerical techniques. After elucidating the actual
behavior of the instantons, we will argue in Sec. 5 that the Great
Divide consists precisely of those potentials which, in the $V_F
\rightarrow 0$ limit, have static domain walls interpolating between
the true and false stationary points of the
potential\footnote{This observation is related to the work of Cvetic
{\it et. al.} on singular domain walls and their relation to CDL
bubbles\cite{cvetic}.};  we also argue that the
Great Divide is appropriately named because its codimension in the space of potentials is one. Finally, in Sec. 6, we will discuss our
results in connection with the picture of eternal inflation put
forward in~\cite{etinf}: we will argue that it is inappropriate to think of potentials describing unstable Minkowski space as having to do with quantum gravity in asymptotically flat space, then discuss what they may, instead, correspond to. A brief summary of our conclusions is given in Sec. 7.

\section{Field equations}
In this paper, we will study a single scalar field with potential of the form
\begin{equation}
V(\phi ) = \mu^4 v\left( \phi /M \right),
\end{equation}
where, defining $x \equiv  \phi /M$, the dimensionless potential $v(x)$ is given by
\begin{equation}\label{potential}
v\left(x\right) =f \left( x \right) -\left(1+z\right) f \left( x_{F} \right),
\end{equation}
where here and henceforth subscripts ``T'' and ``F" will label values at the true and false vacuum, respectively (see Fig.~\ref{vofx}), and where
\begin{equation}
f(x) =\frac{1}{4} x^4 - \frac{b}{3} x^3 -\frac{1}{2} x^2.
\end{equation}
We will tune the parameter $b$ such that the potential has three extrema as shown in Fig.~\ref{vofx},
and has variations of order $1$ between $x_F$ and $x_T$. The non-negative parameter $z$ controls the
false vacuum cosmological constant $V_F$, so that $V_F\rightarrow 0$ as
$z \longrightarrow 0$.\footnote{The way in which we have chosen to tune the vacuum energy is not really appropriate in
many supergravity models.  There, one tunes a constant in the superpotential. If there are excursions in field
space of order $m_P$, this changes the potential in a more complicated way than a simple subtraction. We hope to
return to a study of supergravity models in a future publication.} The general scaling form of the potential is
motivated by considerations of naturalness. Typical potentials which cannot be fit into this form have fine-tuned
dimensionless coefficients and are not stable to radiative corrections.\footnote{The major exception we know of
is the case of moduli in string theory near singular points in moduli space: while the typical potential for moduli
depends on $\phi / m_P$ or $\phi / m_S$, near singular points (where other degrees of freedom become light) the potential can have more rapid variation.}

For many choices of the parameters $b$ and $z$, there will be $0(4)$
invariant instantons, which travel between the basins of attraction
of the minima at $x_{T}$ and $x_{F}$. Together with a scalar field
configuration, $\phi(z)$, the instanton is described by an Euclidean
manifold of the form
\begin{equation}\label{ds2}
ds^2 = dz^2 + \rho^2(z) d\Omega^2,
\end{equation}
where $d\Omega^2$ is the surface element of a unit 3-sphere. Defining the following dimensionless variables:
\begin{equation}
r \equiv \frac{\mu^2 \rho}{M},
\end{equation}
\begin{equation}
s \equiv \frac{\mu^2 z}{M},
\end{equation}
\begin{equation}
\epsilon^2 \equiv \frac{8 \pi M^2}{3 M_{P}^{2}},
\end{equation}
the coupled Euclidean scalar field and Einstein's equations are
\begin{equation}\label{ddotx}
\ddot{x} + \frac{3 \dot{r}}{r}\dot{x} + u' = 0,
\end{equation}
\begin{equation}\label{dotr2}
\dot{r}^2 = 1 + \epsilon^2 r^2 E,
\end{equation}
where $u(x) \equiv -v(x)$, primes and dots, respectively, refer to $x-$ and $s-$ derivatives,
and $E$ is the Euclidean energy of the field, defined as
\begin{equation}
E = \frac{1}{2}\dot{x}^2 + u(x).
\end{equation}
For future reference, the dynamics of the Euclidean energy are determined by the equation
\begin{equation}\label{Edot}
\dot{E} = -3 \frac{\dot{r}}{r} \dot{x}^2.
\end{equation}

When the false vacuum well has positive energy, the Euclidean
spacetime of Eq.~\ref{ds2} is necessarily compact, spanning an interval
between $s=0$ and $s=s_{max}$. To avoid singular solutions to
Eq.~\ref{ddotx}, the field must have zero derivative (i.e. $\dot
x=0$) at $s=0$ and $s=s_{max}$. There will thus be a non-singular
solution to the instanton equations if the boundary conditions
\begin{equation}
r(0)=0, \ \ r(s_{max})=0, \ \ \dot{x}(0) = 0, \ \ \dot{x}(s_{max}) = 0,
\end{equation}
can be met for some set of endpoints in the evolution of $x$ near $x_{T}$ and $x_{F}$.
Solutions with two zeros in $\dot{x}$ will be referred to as as ``single-pass" instantons.
We also note \cite{heretics} that multifield models can be studied using these methods as
well, as long as we restrict attention to instantons for which $\dot{\phi^i} = 0$ only at two points. In that case, however, one might be interested in potentials with more minima and maxima.

The decay rate of the false vacuum is given by
\begin{equation}
\Gamma = A e^{-S_{E}},
\end{equation}
where $A$ is a pre-factor that will be neglected in what follows. The total Euclidean action, $S_{E}$,
is the difference between the action of the instanton, $S_{I}$, (which is negative due to the positive
curvature of the instanton) and the action of the background spacetime, $S_{BG}$ (which is negative and
larger in magnitude than the instanton action)
\begin{equation}
S_{E} = S_{I} - S_{BG}.
\end{equation}
The instanton action is given by
\begin{equation}
S_{I} = -4 \pi^2 \left( \frac{M^4}{\mu^4} \right) \int_{s=0}^{s=s_{max}} ds \left( r^3 u + \frac{r}{\epsilon^2}\right).
\end{equation}
The background subtraction term (for an end-point of the evolution in $x$ near $x_F$) is given by
\begin{equation}\label{SBG}
S_{BG} = \frac{8 \pi^2}{3 \epsilon^4 u_F}.
\end{equation}

In what follows we will be interested in the relative magnitude of the instanton and background actions.
In particular, when the false vacuum cosmological constant is taken to zero, the backgound subtraction
term Eq.~\ref{SBG} diverges. Unless the instanton action scales similarly, the tunneling rate is very
strongly suppressed for small $u_F$.

\section{Approximate analytic solutions}

We can solve Eq.~\ref{ddotx} and \ref{dotr2} exactly when the Euclidean energy remains approximately constant for
a period of time. This can only occur in the neighborhood of the extrema of the potential. The focus of this study
is on transitions from a positive Euclidean energy well at $x_{T}$ to a negative Euclidean energy well at $x_{F}$,
but the results we present below can be used to study arbitrary combinations of positive and  negative energy wells.
The approximate solution to the instanton equations near $x_{H}$ (see Fig.~
\ref{vofx}) was presented in \cite{hackworth}, and is relevant
for the study of oscillating solutions.

Consider the evolution of the field in the neighborhood of $x_{T}$ or $x_{F}$. The field will begin/end with zero
velocity and some displacement, $\delta_{T,F}$, from $x_{T}$ or $x_{F}$. If the variable $\delta_{T,F}$ is small,
then the field will loiter in the neighborhood of the maximum. During this time, the Euclidean energy of the field
will remain roughly constant and, if the velocity remains small, equal to the value of $u$ at the
maximum. Equation~\ref{dotr2}, for the cases of loitering near the true or false vacuum maxima, then reduces to
\begin{equation}
\dot{r}^2 \simeq 1 + \epsilon^2 r^2 u_{T,F},
\label{eq-rloit}
\end{equation}
which can be integrated to yield
\begin{equation}\label{approxr}
r(s) = \frac{1}{\epsilon \sqrt{-u_{T,F}}} \sin \left(\epsilon \sqrt{-u_{T,F}} \right).
\end{equation}
If we take the false vacuum maximum to have $u_{F} < 0$, then we can
recognize this as the metric for Euclidean de Sitter space (the four
sphere). Substituting Eq~\ref{approxr} into Eq.~\ref{ddotx} yields:
\begin{equation}
\ddot{x} + 3 \epsilon \sqrt{-u_{T,F}} \cot\left(\epsilon \sqrt{-u_{T,F}} s\right) \dot{x} + u'(x)=0.
\end{equation}
Since we are trying to find solutions only in the vicinity of the true and false vacuum maxima, we may Taylor
expand the potential about $x_{T,F}$, keeping only the constant and quadratic terms. After making the change of
variables $y=\cos\left(\epsilon \sqrt{-u_{T,F}}\right)$ and $\delta = x - x_{T,F}$, we then obtain
\begin{equation}
\left(1-y^2\right) \frac{d^2 \delta}{d y^2} - 4 y \frac{d \delta}{dy} + \frac{\omega^2}{\epsilon^2 u_{T,F}}  \delta = 0,
\end{equation}
where $\omega^2 \equiv |u_{T,F}''|$. This can be recognized as the hyperspherical differential equation,
the solution of which is given in terms of Legendre functions. After imposing the boundary
conditions $\dot{\delta}(y = 1) = 0$
and $\delta(y=1) = \delta_{T,F}$, we obtain
\begin{equation}\label{legendre}
\delta(y) = \delta_{T,F} \frac{-2 i}{\nu \left(\nu+1\right)} \left(y^2 - 1\right)^{-1/2} P_{\nu}^{1}(y),
\end{equation}
with
\begin{equation}
\nu = - \frac{1}{2} \left(1+\sqrt{9 + \frac{4 \omega^2}{\epsilon^2 u_{T,F}} }   \right).
\end{equation}
For $s \ll \epsilon \sqrt{|u_{T,F}|}$, this solution can be written in terms of Bessel functions.

We have found an approximate analytic solution near the true and false vacuum maxima. However, in order to
construct the entire single-pass instanton we must evolve across regions of the potential in which our
approximations break down. This requires a numerical approach, which will be presented in Sec.~\ref{numres}.

\section{The $V_{F} \rightarrow 0$ limit}\label{Vf0limit}

We are now in a position to re-examine some of the conclusions of \cite{etinf}. Two of the authors (TB and MJ)
conjectured that for all $\epsilon$ the instanton describing a transition from a positive energy false vacuum
to a negative energy true vacuum approaches a finite size as $z \rightarrow 0$, and therefore the instanton action
would not scale with the background subtraction term. We argued (to ourselves) that there would also be a  flat space
instanton which existed for $z = 0$, by a version of Coleman's overshoot/undershoot argument. This implied a
discontinuous limit as the false vacuum energy was sent to zero.

Here, we will present numerical and analytical arguments that below some (potential dependent) $\epsilon_c$
there are in fact large dS instantons that asymptote as $z \rightarrow 0$ to the flat space instanton.
Above $\epsilon_c$, there are finite-size instantons with finite action as $z \rightarrow 0$, but no
flat space instanton. At $\epsilon_{c}$ (on the Great Divide), we will find that the instanton for $z=0$ is
a static domain wall solution of the coupled Euclidean Einstein and field equations.

\subsection{Small $\epsilon$}
\label{smalle}
Let us explore the small $\epsilon$ case first, and argue that if a single-pass
instanton exists, it {\it must} resemble the dimensionless de Sitter metric, Eq.~\ref{approxr},
over most of its volume. From Eq.~\ref{dotr2}, we see that the Euclidean energy, which is bounded
from below by the value $u(x_{H})$ of the potential at the Hawking-Moss maximum, must be negative
for a turn-around in $r$ to occur. If there is a turn-around, the value of $r$ at this point, $r_m$, will be
\begin{equation}
r_{m} = \frac{1}{\epsilon \sqrt{-E_m}}.
\end{equation}
Since the Euclidean energy is bounded, as $\epsilon$ is decreased, $r_{m}$ must increase. If there is a compact
nonsingular instanton, the field must evolve in such a way to facilitate this growth in $r$. When the field is not
in the vicinity of the extrema of the potential, it will move between the potential wells in a time of order one.
During this time, $r$ will grow to some $\epsilon$ independent size. Thus, for $r$ to become large enough to find
a turn-around in the small $\epsilon$ limit, the field must loiter in the vicinity of one of the extrema of the potential.

Loitering near the Hawking-Moss maximum leads to an oscillatory
motion, because this is a minimum of the Euclidean potential. There
are non-singular solutions which make of order ${1\over\epsilon}$
oscillations before ending up in the basin of $x_F$. These are not
single pass instantons. Loitering near the true vacuum maximum will
cause $r$ to grow as in Eq.~\ref{approxr} (linearly if $s \ll
\epsilon \sqrt{u_T}$). However, because the friction term decays
during the loitering phase, these solutions will in general have too
much energy and overshoot the false vacuum maximum. For intermediate
values of $\epsilon$,the growth in $r$ near the true vacuum becomes
important, as we will see below.

The only viable option is then that the field be near $x_{F}$ at the turn-around in $r$.
If we take the end-point near $x_{F}$ to be at $s=0$, the field must remain near $x_{F}$
until $r=r_{m}$. This evolution should be well described by the analytic solution Eq.~\ref{legendre}
derived in the previous section. The Euclidean energy at $r_{m}$ will be given by
\begin{equation}
E_m \simeq u_{F} + \frac{1}{2} \dot{\delta}_{m}^{2} - \frac{\omega^2}{2}\delta_{m}^{2}.
\end{equation}
We can write $\delta_{m}$  and $\dot{\delta}_{m}$ in terms of Gamma functions
\begin{equation}\label{gammad}
\delta_{m} = \delta(s=\pi/2 \epsilon \sqrt{v_{F}}) = \delta_{F} \frac{\sqrt{\pi}}{\Gamma\left(1-\frac{\nu}{2} \right) \Gamma\left(\frac{3}{2} + \frac{\nu}{2} \right)},
\end{equation}
and
\begin{equation}\label{gammadot}
\dot{\delta}_{m} = \dot{\delta}(s=\pi/2 \epsilon \sqrt{v_{F}}) = -\delta_{F} \epsilon \sqrt{\pi v_{F}} \frac{2+\nu}{\Gamma\left(\frac{1}{2}-\frac{\nu}{2} \right) \Gamma\left(2+\frac{\nu}{2}\right)}.
\end{equation}
This limit will be an important component of the numerical
scheme presented in the following section. We note that $\delta_m$
and $\dot{\delta}_m$ are of the same order of magnitude, and must be
much smaller in magnitude than $v_{F}$ for our approximation scheme
to remain self-consistent. This can always be arranged by making
$\delta_F$ of order $\exp(\frac{-1}{\epsilon \sqrt{v_{F}}})$. Thus,
we can see that there is a self-consistent solution in the vicinity
of $x_{F}$ which tracks the de Sitter solution until $r_{m}$.

In fact, it is necessary, for small $\epsilon$, to choose $\delta_F$
small enough that the de Sitter/Legendre approximation remains valid
until $s = {\pi\over {\epsilon\sqrt{v_F}}} - o(1) $. If we do not do
this, then $x(s)$ moves rapidly away from $x_F$ on a time scale of
$o(1)$, while $r(s)$ is still $\gg 1$.  It will either overshoot
$x_T$ or stop and fall back, long before the second zero of $r(s)$
is reached.  In neither case do we get a single pass instanton. The
rest of the instanton consists of a traverse from the vicinity of
the false vacuum, to the basin of attraction of the true vacuum, in
a time of $o(1)$ ($\epsilon$-independent for small $\epsilon$).  It is important that, since $r \ll 1/\epsilon$  during this traverse, Eq.~\ref{eq-rloit} indicates that $r(s)$ is approximately linear in this period, and indeed also linear for a long period before $x(s)$ leaves the vicinity of the false vacuum.

It is convenient to think of the rest of the instanton as a
function of a new time variable $t$ which starts at $t=0$ near the
true vacuum and increases toward the false vacuum so that
$d/dt\equiv-d/ds$. Since $r(t) \approx t$ when $r \ll 1/\epsilon$, we have
\begin{equation}\label{colemaneq}
\frac{d^2x}{dt^2} + {3\over t} \frac{dx}{dt} = -u^{\prime } (x),
\end{equation}
with the boundary conditions $\frac{dx}{dt} (t = 0) = 0$ and $x_{H} < x(t=0) < x_T$.

This equation is just the equation for an instanton in quantum field
theory, neglecting gravitational effects. Coleman~\cite{coleman}
showed that one can find solutions which start in the basin of
attraction of the true minimum, and get arbitrarily close to (or
even overshoot) the false minimum.  Eq.~\ref{colemaneq} is $\epsilon$-independent, but
as $\epsilon$ goes to zero, the range of $t$ over which it is a good approximation to the real instanton solution grows as $1/\epsilon$.  Thus,
for small enough $\epsilon$, we can use Coleman's argument to show
that there are solutions of Eq.~\ref{colemaneq}, which are
non-singular at $t = 0$ and penetrate into the region where the
Legendre approximation is valid. By varying the initial position
$x(t = 0)$ among all such solutions, we can tune the logarithmic
derivative of $x$ at a given point $t^*$ where both approximations
are valid, within a finite range.

The conditions that the two solutions match at some point $(t^*,
s^*)$ are
\begin{equation}\label{match1}
t^* = {1\over {\epsilon\sqrt{v_F}}} \sin (\epsilon \sqrt{v_F}
s^*),
\end{equation}
\begin{equation}\label{match2}
\frac{1}{x(s^*)}\frac{dx}{ds} = - \frac{1}{x(t^*)} \frac{dx}{dt} ,
\end{equation}
\begin{equation}\label{match3}
x(s^* ) = x(t^*),
\end{equation}
where functions of $s^*$ are in the de Sitter/Legendre approximation
and functions of $t^*$ are in the zero-gravity approximation. Once
we know that there is a range of $x(t=0)$ for which $x(t)$
penetrates into the range where the Legendre approximation is valid,
we can tune $x(s=0)$ to satisfy the last condition. We know that $s*$ is large for very small $\epsilon$, of order ${\pi\over {\epsilon\sqrt{v_F}}} - o(1)$, in which case the first condition becomes $t* = s*$. 

$x(t=0)$ is then tuned to match the logarithmic derivatives.
Although there is a range of $s$ over which $x(s)$ is rapidly
varying, its logarithmic derivative is roughly constant over that
range. The only place where the logarithmic derivative is large, is
near the second zero of the sine, but for small $\epsilon$ the
matching occurs far from that region ($t^*$ large but $\ll
{1\over{\epsilon\sqrt{v_F}}} $). It is thus plausible that by
varying $s^*$ and $x(t = 0)$ we can satisfy both of
Equations~\ref{match1} and~\ref{match2}.  If this is the case, then
 a non-singular, large radius instanton exists. As $v_F \rightarrow 0$,
this goes over smoothly to an \lq\lq instanton for the decay of
asymptotically flat space".

The argument above indicates the possibility of a true asymptotic
matching of solutions of the non-gravitational equations to
solutions of the de Sitter/Legendre approximation over a range of
$s$ which grows as $\epsilon\rightarrow 0$.  Since we cannot exhibit
solutions of the non-gravitational equations exactly, our argument
is not completely rigorous. In the next section we will present
numerical calculations, which show that it is correct.

\subsection{Numerical results for small $\epsilon$}\label{numres}

To confirm the validity of the conclusions above, we have undertaken a semi-analytic
search for single pass instantons in a potential with a positive false vacuum and a
negative true vacuum. Here, we will focus on the potential shown in Fig.~\ref{potentialdiag},
though qualitatively our results are potential independent (we have confirmed this by studying a variety of potentials).

\EPSFIGURE{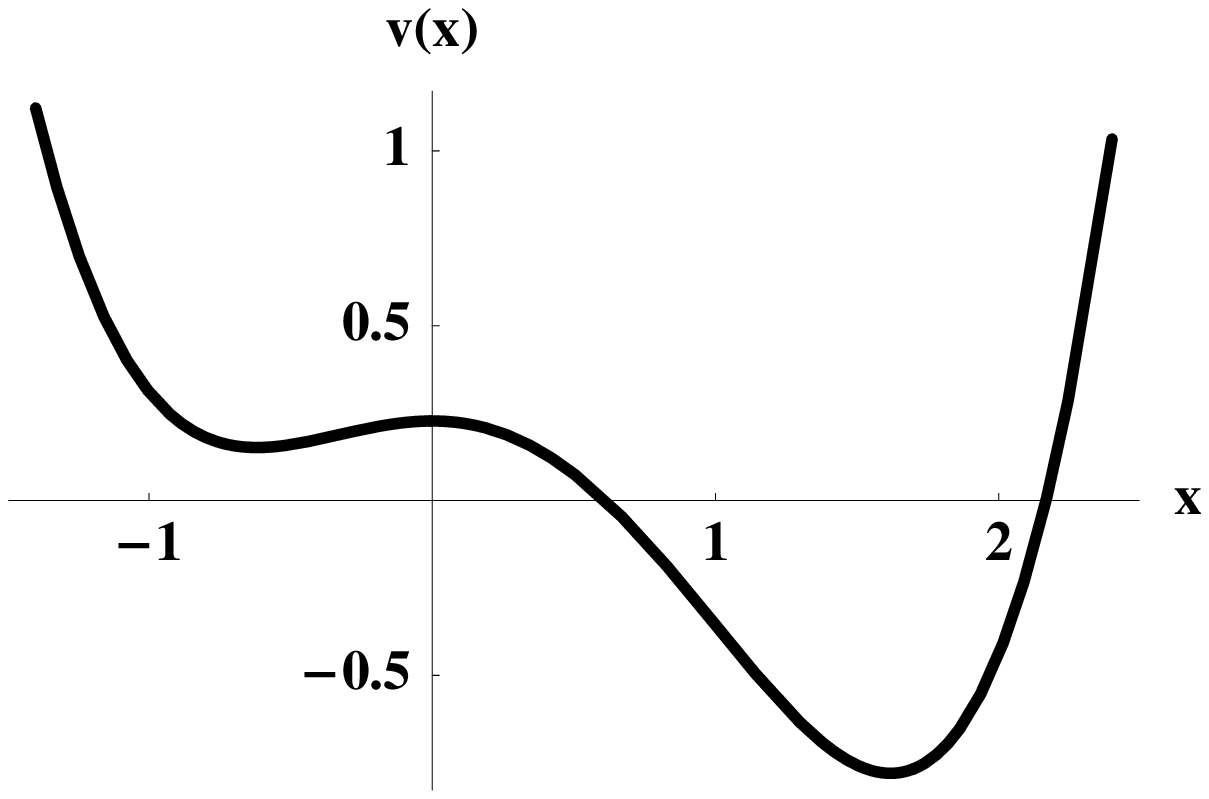,width=6cm}{The potential, $v(x)$, used for the numerics. The parameter $b$ is fixed at $b=1$, and $z$ will be allowed to vary (this plot shows $z=1$).\label{potentialdiag}}

The strategy is to use the matching scheme discussed in
Section~\ref{smalle}. We will relax the zero-gravity approximation
for the evolution from the true vacuum well to the false vacuum
well, and numerically evolve Eqs.~\ref{dotr2} and~\ref{ddotx}. To fix
the initial conditions of the numerical evolution from the true
vacuum side of the potential, we will use an analytic solution to
evolve for the first time step. If it is near $x_T$, we use
Eq.~\ref{legendre}; if not, we approximate the potential as linear, yielding a
$\delta(s) \propto s^2$. We then evolve and attempt to match
onto the de Sitter/Legendre approximation (Eq.~\ref{approxr}
and~\ref{legendre}) when the field approaches $x_{F}$. Of course, we
are not guaranteed to find a match for all $\epsilon$. It was shown
by Coleman and De Luccia \cite{cdl}, that in the thin-wall limit
there are cases where the transition from a positive (Euclidean)
energy well to a zero energy well is forbidden. This occurs when the
positive energy at the true vacuum maximum becomes too small, so that an
over-shoot solution becomes impossible. This would prevent the
instanton from ever entering a regime where the de Sitter/Legendre
approximation was valid.

The need for a semi-analytic approach is evident from the fantastically small
displacement from the false vacuum required to find solutions with large $r_m$. Numerically evolving the solution
over the entire trajectory would become impossible as the field approaches $x_F$. Also for reasons of numerical tractability, we match the solutions at
 at $r_{m}$, where $s=\pi/(2 \epsilon \sqrt{v_{F}})$, and the Legendre function can be written in terms
of (calculable) $\Gamma-$ functions as in Eq.~\ref{gammad} and~\ref{gammadot}.

This method also has its limitations. For small enough $\epsilon
\sqrt{v_{F}}$, we may be trying to compare field velocities at a precision that is not achievable by the
numerical integrator. Despite these difficulties, we have been able
to construct a number of instantons in the intermediate $\epsilon$
regime, examples of which are shown in Fig.~\ref{biginst}. It can be
seen in this plots that as $z \rightarrow 0$, these instantons are
growing.  Since we have shown that a matching is possible at
$r_{m}$, as $v_{F} \rightarrow 0$, by the argument given in Sec.~\ref{smalle}, these instantons must scale with the background subtraction term.

\EPSFIGURE{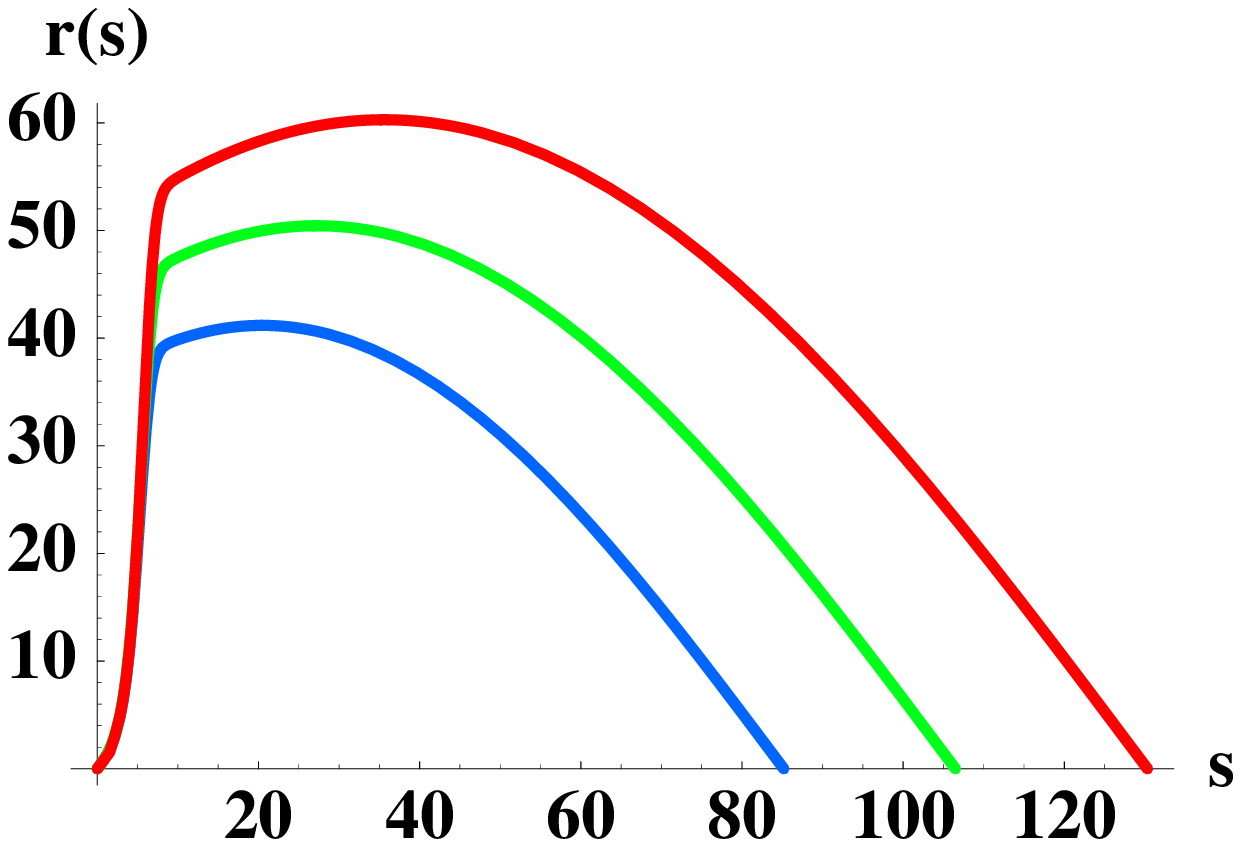,width=12cm}{Evolution of $r(s)$ for $\epsilon=.72$ and $z=(.01,.008,.006)$ from bottom to top. The matching between the analytic and numeric solutions occurs at the maximum of $r$, $r_m$.\label{biginst}}

\subsection{Large $\epsilon$}\label{largee}
To study large\footnote{By large we mean of order one. While the formalism will  accommodate arbitrarily large values of $\epsilon$, there will be an $\epsilon$ after which only the Hawking-Moss instanton exists.} values of $\epsilon$, where the approximations introduced above are not necessarily valid,
we must take an entirely numerical approach.  We choose to begin the evolution from the true vacuum side
of the potential, varying $\delta_{T}$ until a solution is found. To fix the initial conditions of the numerics,
we will again use an analytic solution to evolve for the first time step as described in the previous section.

Shown in Fig.~\ref{smallinstx} is the evolution in $x$ for $\epsilon=.85$ as $z \rightarrow 0$. Shown in Fig.~\ref{smallinstr} is the evolution in $r$ with the same parameters. It can be seen that as $z \rightarrow 0$, the instanton approaches a constant, finite size. Therefore, for large $\epsilon$, the instanton action will not scale with the background subtraction term.

To discuss the continuity of the limit $V_{F} \rightarrow 0$, we must first determine in which cases there is an
instanton for $V_{F} = 0$. If this instanton describes the decay of a spacetime with exactly zero cosmological constant,
then the evolution in $r$ must be from $r(s=0)=0$ to $r(s=\infty)=\infty$. The field will be moving from some initial
position near $x_{T}$ at $s=0$ to {\em exactly} $x_{F}$ at $s=\infty$.  If, starting near $x_T$,  there is a region of $\delta_T$-space in which over-shoot occurs, then there must be a second zero in $\dot{x}$. The question is then what value $r$ takes at the second zero of $\dot{x}$.

In all of the numerical examples we have studied with $z=0$, we find that $r=0$ at the second zero of $\dot{x}$.
The turn-around in $r$ in these cases is not caused by loitering in the vicinity of a negative energy extremum of
the potential. Instead, as the field is climbing towards $x_{F}$, the negative potential energy comes to dominate over
the kinetic energy. Since $\epsilon$ is rather large, $r$ does not need to grow very large to cause a turn-around in r.
Since the end-points of this instanton are on the boundaries of the unique over- and under-shoot regions of the potential,
there is no other single-pass instanton with $r(s=\infty)=\infty$.

\EPSFIGURE{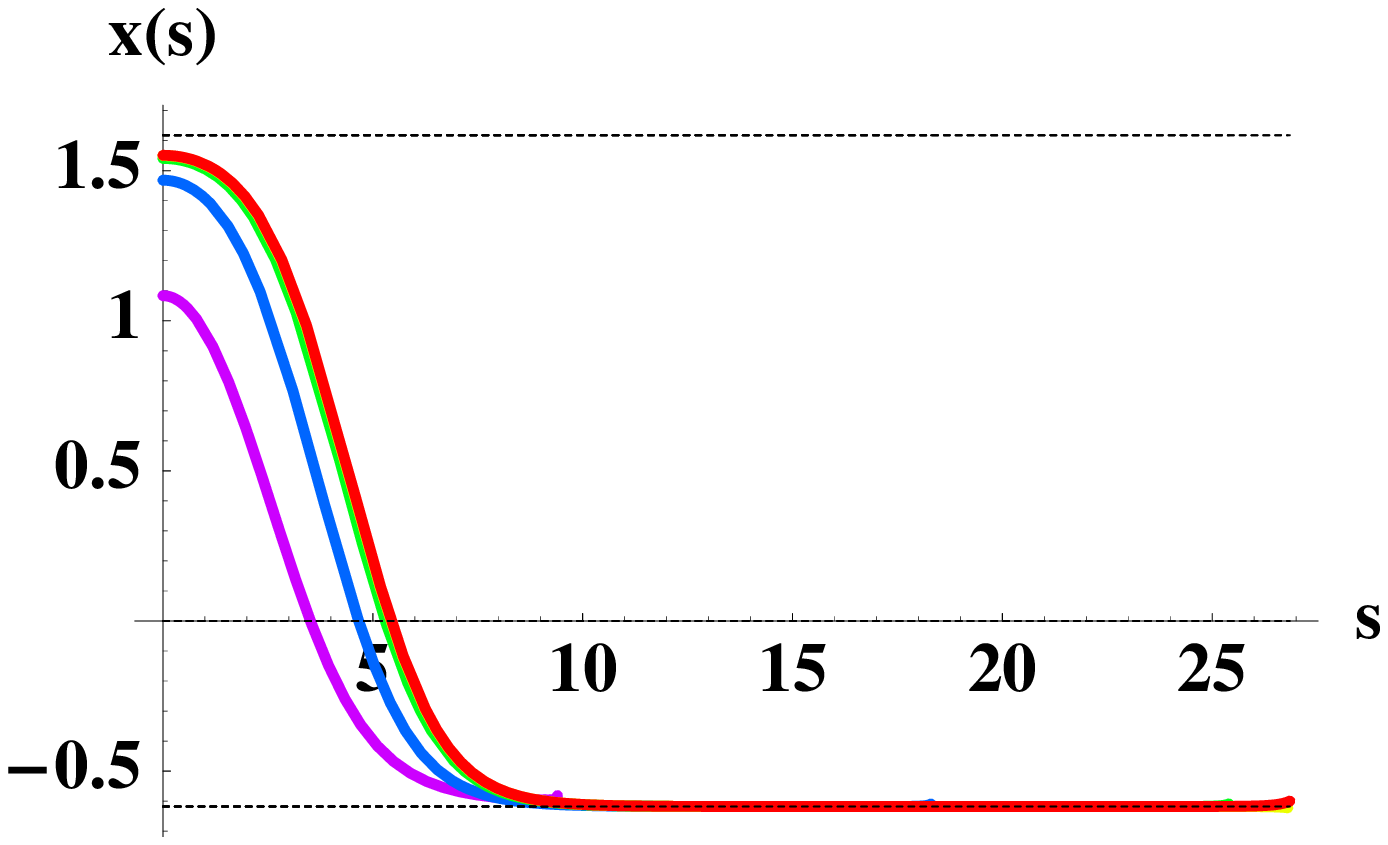,width=12cm}{The evolution of $x(s)$ for $\epsilon=.85$ and $z=(1, .1, .01, .001, .0001)$ from bottom to top. The dashed horizontal lines indicate the positions $x_T$ (top) and $x_F$ (bottom) . \label{smallinstx}}

\EPSFIGURE{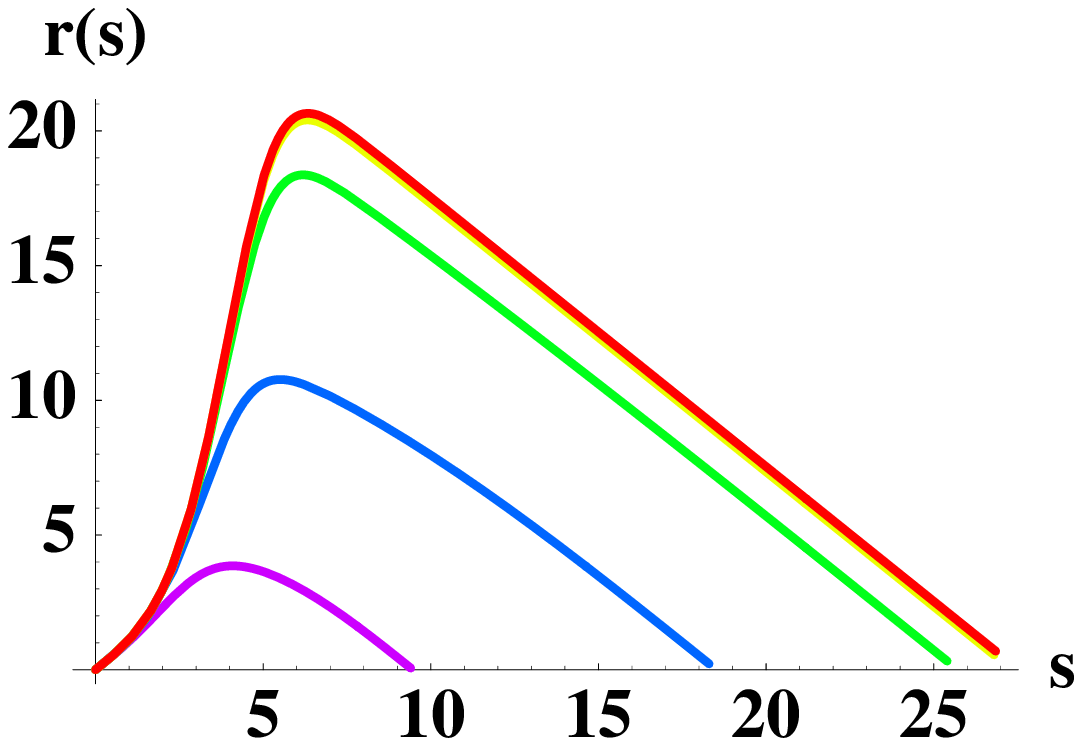,width=12cm}{The evolution of $r(s)$ for $\epsilon=.85$ and $z=(1, .1, .01, .001, .0001)$ from bottom to top. \label{smallinstr}}

\section{The Great Divide}
In this section we show that, for any potential $v(x)$, there is a
critical value of $\epsilon$ for which planar domain wall solutions
exist. As one goes from the small to the large
$\epsilon$ regime, there is a transition point between the two
behaviors discussed in Section~\ref{Vf0limit}. We will define
$\epsilon_{c}$ as the transition point in the case where $z=0$ (when
the false vacuum well has zero energy).

We have found instantons (with $z=0$) for a variety of $\epsilon$ near $\epsilon_{c}$ as shown in Fig.~\ref{instseries}. The evolution of the field is from the vicinity of $x_T$ at $s=0$ to $x_F$ at $s=\infty$. Of course, we cannot track the entire evolution, but we can follow it for some finite time scale by tuning $\delta_T$ to approach the boundary between the under- and over-shoot solutions. It can be seen from these numerical examples that $r$ is growing very large in the vicinity of the true vacuum.

As we approach $\epsilon_{c}$, the initial displacement on the true vacuum side, $\delta_{T}$, is decreasing as shown in Fig.~\ref{x0plot}. Because
we are starting with more energy on the true vacuum side of the
potential, we must send $\delta_{F} \rightarrow 0$ as well.
Therefore, at this critical value of $\epsilon$, the instanton
interpolates exactly between $x_{T}$ at $s=-\infty$ and $x_{F}$ at
$s=+\infty$. Also, note that after we analytically continue to the
Lorentzian solution, the interior of the CDL bubble will be
infinitely large. This solution therefore describes a static domain
wall.

\EPSFIGURE{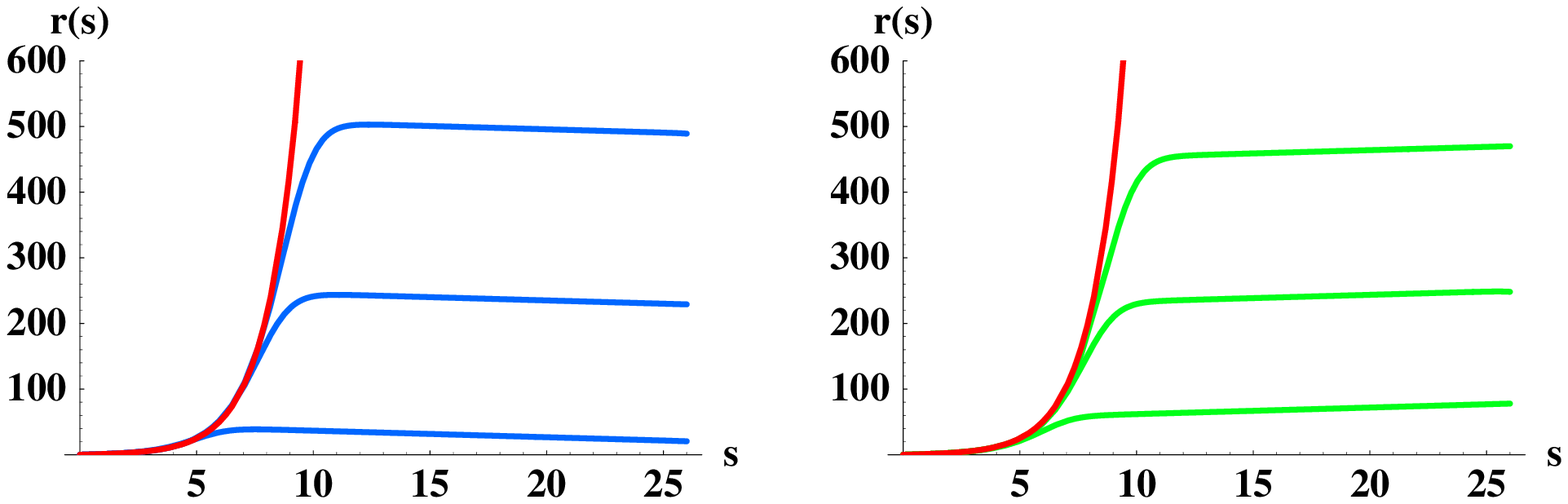,width=16cm}{The evolution of $r(s)$ for $z=0$ on either side of $\epsilon_c$. Shown on the left are values of $\epsilon > \epsilon_c$ in blue ($\epsilon = (.8, .75, .745)$ from bottom to top) and $\epsilon_C \sim .74$ in red. The instantons with $\epsilon > \epsilon_c$ are compact, having two zeros in $r$.  On the right are values of $\epsilon < \epsilon_c$ ($\epsilon = (.7, .73, .735)$ from bottom to top) in green and  and  $\epsilon_c$ in red. The instantons with $\epsilon < \epsilon_c$ are not compact, with $r \rightarrow \infty$ as $s \rightarrow \infty$. \label{instseries}}

\EPSFIGURE{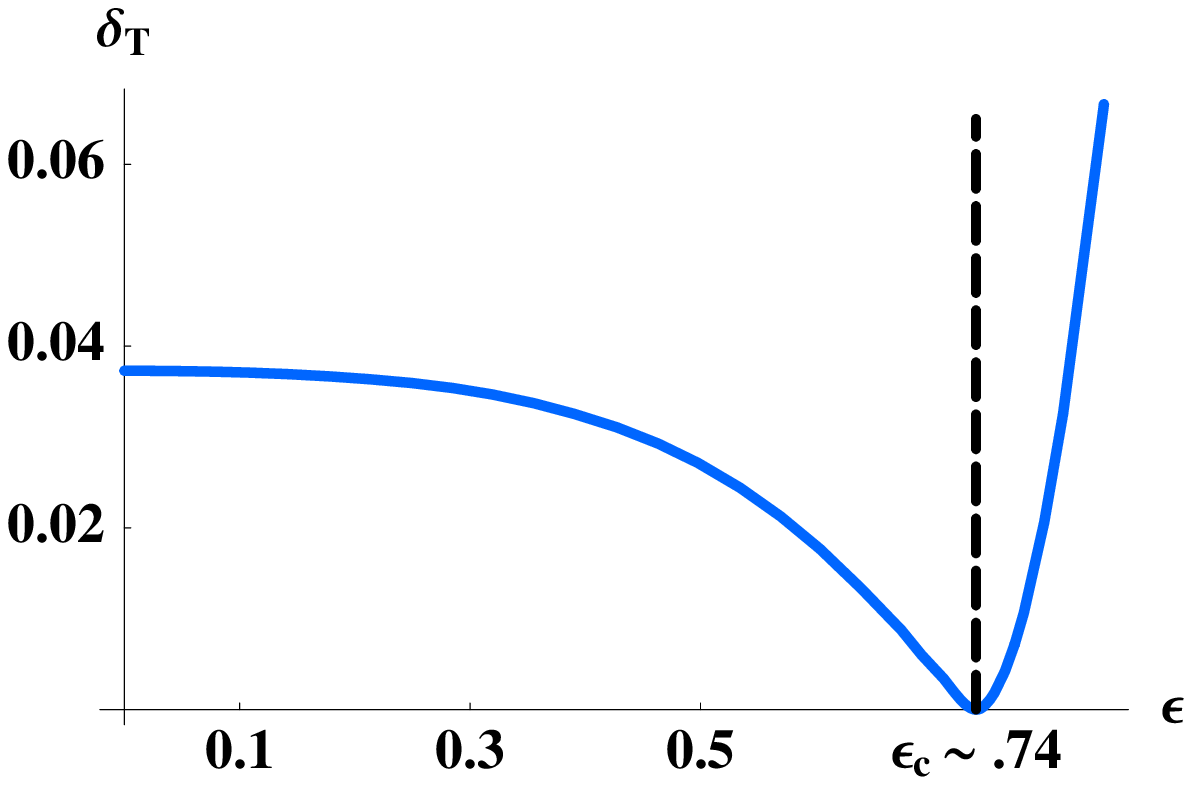,width=12cm}{It can be seen in this plot of $\delta_{T}$ vs $\epsilon$ for the case where $z=0$ that there is an $\epsilon_{c}$ for which $\delta_{T} \rightarrow 0$. Below this value, $\delta_{T}$ is approaching the zero-gravity solution, and above it, $\delta_{T} \rightarrow x_T-x_H$. \label{x0plot}}

We can understand this behavior by looking at the energetics of the
evolution from $x_{T}$ to $x_{F}$. The instanton equations in the
critical limit approach the static domain wall equations
\begin{equation}\label{ddotxdw}
\ddot{x} + \frac{3 \dot{r}}{r}\dot{x} + u' = 0,
\end{equation}
\begin{equation}\label{dotr2dw}
\dot{r}^2 =  \epsilon^2 r^2 E,
\end{equation}
$s$ now runs between $-\infty$ and $\infty$, and a domain wall
solution asymptotes to the two vacua on opposite sides.
 The energy is always decreasing along the
trajectory from the true to the false vacuum well. The question is
whether $x$ can lose just enough energy during its traverse to
asymptote to $x_F$ without overshooting. If $\epsilon =0$ the answer
is clearly no, because energy is conserved. The solution overshoots
the false vacuum. This persists for very small $\epsilon$. On the
other hand, in the mathematical limit $ \epsilon \gg 1$, the
friction term dominates the motion and $x$ undershoots in a finite
time. It follows that there is a critical value of $\epsilon$ where
$x$ indeed asymptotes to $x_T$ and we have a static domain wall
solution in the presence of gravity. The critical value is clearly
$o(1)$. Since we have found such a solution by tuning a single
parameter, the codimension of the subset of potentials which have a
domain wall is $1$, and the subset forms a Great Divide in the space
of potentials.

We have shown both that there is a critical value of $\epsilon$ at
which domain walls exists, and that the flat space instanton
solution, which exists below the Divide, approaches the domain wall
solution at this critical value.   Above the divide, the flat space
instanton and the associated large instantons for small $v_F$,
disappear. Flat space is stable, and the stability of nearly flat
dS spaces has a clear entropic explanation.

\section{\bf Below the great divide}

In~\cite{etinf}, along with the conjecture of a discontinuity of the tunneling action at $V_F\rightarrow 0$ came
 a (retrospectively flawed) physical argument to explain the discontinuity, based on the physical picture of quantized
 dS space adumbrated in \cite{tbds}.  In that picture, quantized dS space is equipped with two operators: the
 static Hamiltonian $H$, and the Poincare Hamiltonian $P_0$; these
satisfy a finite-dimensional approximation to the commutation relation
\begin{equation}
[H, P_0 ] \sim {1\over R} P_0 ,
\end{equation}
where $R$ is the de Sitter radius. The eigenvalues of $H$ are highly degenerate, and bounded by something of order
the dS temperature, $T_{dS} = {1\over {2\pi R}}$.   The low-lying eigenstates of $P_0$ are metastable (when evolved using
$H$), and correspond to states localized in a given horizon volume; the lowest lying eigenstates have small degeneracies,
and the ground state is unique.  The conjectured discontinuity in the tunneling probability was alleged to be related to
the fact that the for finite $V_F$ the CDL instanton describes the decay of the thermal ensemble of $H$ eigenstates (a system of high-entropy), but that for vanishing $V_F$ it describes the decay of a low-entropy system consisting just
of the single $P_0$ ground state.

The flaw in this argument is that it hypothesizes both a stable $P_0$
eigenstate, and also the decay of that stable system. That is, the existence of
the CDL instanton for potentials below the great divide is, in fact,
evidence that these low energy effective theories do not correspond
to limits of theories describing asymptotically flat space-time.

The conformal boundary of the Lorentzian continuation of the CDL instanton is not the same as that
of Minkowski space: in the usual parametrization $(u,\Omega )$ of future null infinity, ${\cal I}^+$, in terms of a
null coordinate $u$ and a transverse sphere, the boundary becomes geodesically incomplete because the asymptotic bubble
wall hits ${\cal I}^+$ at a finite value of $u$. Neither the Lorentz group (consisting of the conformal group of the
sphere accompanied by a rescaling of $u$) nor the time translation group (the generator of which is
just $P_0 = {\partial\over \partial
u}$, in a particular Lorentz frame) is an asymptotic symmetry of this spacetime.  Thus, the
``explanation" of an hypothetical discontinuity in~\cite{etinf} was
based on an equally hypothetical operator $P_0$.  Neither exists.

If potentials below the Great Divide do not correspond to effective theories of gravity in asymptotically
 flat space, what do they correspond to? Two possibilities consistent with the authors' current understanding of quantum gravity are:
\begin{enumerate}

\item Nothing.  That is, there simply are no theories of quantum gravity which give
rise to such potentials.

\item These theories correspond to models of quantum gravity which, in the $V_F
\rightarrow 0$ limit under consideration, actually contain only a finite number of excitations of
the  Minkowski solution. This would remove the apparent contradiction between the infinite number
of states of the would-be asymptotically flat space and the finitely bounded entropy of the maximal-area
causal diamond in the Big Crunch.

\end{enumerate}

The confusion may be simplified enormously if the
conjecture of \cite{tbfolly} is accepted. According to that
hypotheses, the only viable quantum theories of asymptotically flat
space time are exactly supersymmetric, and all models with a vacuum
energy that can be tuned to be arbitrarily small become exactly
supersymmetric in that limit. At the moment, this conjecture is
valid for all models which have been derived from string theory in a
reliable manner. The whole concept of the Great Divide is defined in
terms of one-parameter families of potentials, with vacuum energy
that can be tuned to zero. The conjecture of \cite{tbfolly} thus
implies that all valid models of quantum gravity will fall above the
Great Divide; which is hypothesis $1$ above.

\section{Conclusions}

We have seen that there is a rich variety of behaviors of instantons
describing the transition from positive or zero energy false vacuum
to a negative energy Big Crunch. The complete picture is more
detailed than was conjectured in \cite{etinf}, and different
than the conventional (thin-wall) wisdom suggests. For small values
of $\epsilon$, we have shown that there {\em does} exist an
instanton which resembles Euclidean de Sitter over most of its
volume. As the false vacuum energy is taken to zero, the instanton
action scales with the background subtraction, and there is no
discontinuity in the tunneling rate. However, the analytically
continued bubble wall removes a section of the conformal boundary of
Minkowski space, providing evidence that low energy effective
theories with small $\epsilon$ do not correspond to limits of
theories describing truly asymptotically flat space-time.

We have found that there exists a static domain wall solution at a critical value of $\epsilon$ ($\epsilon_c$).
The critical value of $\epsilon$ corresponds to a Great Divide in the space of potentials, of codimension one.
Below $\epsilon_c$, we find the behavior described in the previous paragraph. Above this value of $\epsilon$,
we find compact instantons which {\em do not} resemble Euclidean de Sitter. The instanton action approaches a
constant as the false vacuum energy goes to zero, but the discontinuity claimed in \cite{etinf} does not exist.
We find that there is no non-compact instanton describing the decay of the zero-energy false vacuum, and therefore
as the false vacuum energy is decreased, the diverging background subtraction will cause an infinite suppression
of the tunneling rate.

In \cite{etinf}, two of the authors proposed a regulated model of
eternal inflation for potential landscapes with only non-vanishing
vacuum energies. According to that model the system has a finite
number of quantum states, and for most of its time evolution it
resembles the dS space of lowest positive vacuum energy\footnote{If the minimum with lowest absolute value of the
vacuum energy is negative, then this statement might be corrected to "for
most of the period during which local observers exist it resembles
the dS space of lowest positive vacuum energy".}. This model remains valid
for potentials above the Great Divide. For such potentials,
tunneling amplitudes out of dS space are suppressed in a way which
is attributable to the principle of detailed balance, and entropic
effects.

The other observation of \cite{etinf} which remains unchanged by our
new results is the remark that metastable SUSY violating vacua of
flat space field theories can be viable models of the real world,
within the context of Cosmological SUSY Breaking.   That is, if we
assume that the vacuum energy is tunable and that the limit of vanishing vacuum energy is a supersymmetric theory in asymptotically flat space, then we are
above the Great Divide.  For finite $\Lambda$ the probability for
the meta-stable vacuum to make a transition to a Big Crunch is of
order $e^{-\pi (RM_P )^2 }$. This is not a decay, and it has no
phenomenological relevance.

Our new results raise interesting questions about the interpretation
of models below the Great Divide. The study of these models will be the subject of a future paper.

\section{Acknowledgments}
We would like to thank R. Bousso and B. Freivogel for stimulating
discussions.   This work was supported in part by the Department of
Energy under grant number DE-FG03-92ER40689.

%%%%%%%%%%%%%%%%%%%%%%%%%%%%%%%%%%%%%%%%%%%%%%%%%%%%%%%%%%%%%%%%%%%%%%%%%=

%%%

%                      REFERENCES                                        =

  %

%%%%%%%%%%%%%%%%%%%%%%%%%%%%%%%%%%%%%%%%%%%%%%%%%%%%%%%%%%%%%%%%%%%%%%%%%=

%%%

%\newpage

\end{document}